\documentclass[conference]{IEEEtran}
\IEEEoverridecommandlockouts
\usepackage{cite}
\usepackage{amsmath,amssymb,amsfonts}
\usepackage{algorithmic}
\usepackage{graphicx}
\usepackage{textcomp}
\usepackage{array}
\usepackage{xcolor}
\usepackage{multirow}
\usepackage{url}
\usepackage{booktabs}
\usepackage{multirow}

\usepackage{booktabs}
\usepackage{multirow}
    
\def\BibTeX{{\rm B\kern-.05em{\sc i\kern-.025em b}\kern-.08em
    T\kern-.1667em\lower.7ex\hbox{E}\kern-.125emX}}
    
\makeatletter
    \newcommand{\linebreakand}{%
      \end{@IEEEauthorhalign}
      \hfill\mbox{}\par
      \mbox{}\hfill\begin{@IEEEauthorhalign}
    }
\makeatother
    
\begin{document}

\title{ Real-Time EMG Signal Classification via Recurrent Neural Networks}

\author{
\IEEEauthorblockN{Reza Bagherian Azhiri }
\IEEEauthorblockA{\textit{Predictive Analytics and Technologies Lab, ME Dept.} \\
\textit{The University of Texas at Dallas}\\
Richardson, TX, USA \\
reza.azhiri@utdallas.edu}
\and
\IEEEauthorblockN{Mohammad Esmaeili}
\IEEEauthorblockA{\textit{Department of Electrical and Computer Engineering} \\
\textit{The University of Texas at Dallas}\\
Richardson, TX, USA \\
esmaeili@utdallas.edu}
\linebreakand
\IEEEauthorblockN{Mehrdad Nourani}
\IEEEauthorblockA{\textit{Predictive Analytics and Technologies Lab, ECE Dept.} \\
\textit{The University of Texas at Dallas}\\
Richardson, TX, USA \\
nourani@utdallas.edu}
}
\maketitle

\begin{abstract}
Real-time classification of Electromyography signals is the most challenging part of controlling a prosthetic hand. Achieving a high classification accuracy of EMG signals in a short delay time is still challenging. Recurrent neural networks (RNNs) are artificial neural network architectures that are appropriate for sequential data such as EMG. In this paper, after extracting features from a hybrid time-frequency domain (discrete Wavelet transform), we utilize a set of recurrent neural network-based architectures to increase the classification accuracy and reduce the prediction delay time. The performances of these architectures are compared and in general outperform other state-of-the-art methods by achieving $96\%$ classification accuracy in $600$ msec. 
\end{abstract}

\begin{IEEEkeywords}
Electromyography, Recurrent Neural Network, Real-Time classification, Deep Learning, Wavelet Transform.
\end{IEEEkeywords}

\section{Introduction}
Electromyography (EMG)-based pattern recognition of finger movements has been widely accepted by researchers as a promising method for controlling of prosthetic hands. The EMG signal is the result of action potential of muscle tissues after contraction. This transmitted electrical signal is captured by skin-mounted sensors placed on the target muscle. Due to monotonic relation between finger movement and that of its associated EMG signal, the pattern recognition of hand gesture could be detected. This pattern recognition is examined in either offline or online mode. In offline mode, the major goal is to achieve a higher accuracy~\cite{rouhani2018algorithms}. The extracted features from the EMG signal and utilized classifier both impact the accuracy. Generally speaking, three major feature sets are utilized for EMG signals classification: in time domain (TD), frequency domain (FD) and time-frequency domain. 
Various optimization methods such as Particle Swarm Optimization (PSO) and Genetic algorithm (GA)~\cite{2015novel, Lima2018genetic} can be employed to select a set of features with significant importance.
Also, a plethora number of classifiers have been reported by researchers including decision trees~\cite{Espinoza_DT_SVM}, random forest~\cite{Li_RF}, K nearest neighbor (KNN)~\cite{Nazarpour_KNN, ref}, naïve Bayes classifier~\cite{Praveen_naive}, multilayer perceptron (MLP)~\cite{jaramillo2017real}, gradient boosting (GB)~\cite{bhattachargee2019finger}, support vector machine (SVM)~\cite{Heydarzade, gailey_svm} and extreme value machine (EVM)~\cite{azhiri2021emg-based}.

\begin{figure*}[h]
\centering
\includegraphics[width=0.99\textwidth]{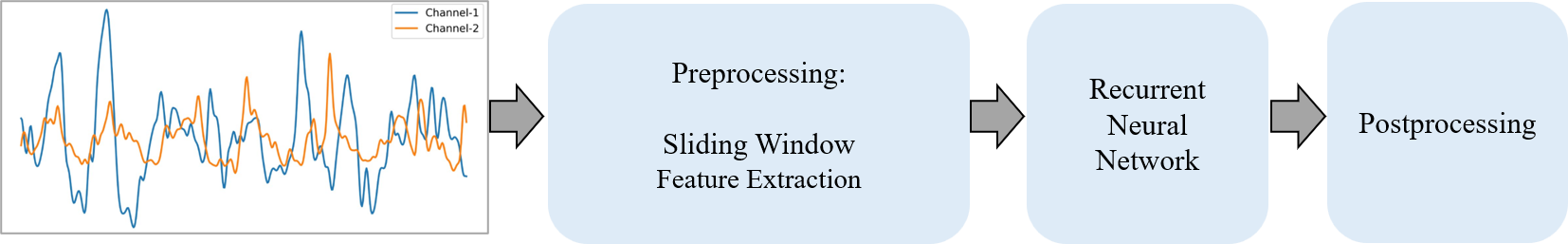}
\caption{The overall procedure of our proposed model.}
\label{Teaser}
\end{figure*}
 
Esa et al.~\cite{esa2018electromyography} extracted Hudgins features, root mean square (RMS) and combination of all these features and then employed SVM as a classifier to get the accuracy of $86.67\%$ on index finger and $96.67\%$ on thumb finger. 
Reference~\cite{Heydarzade} performed spectral analysis on EMG signals to extract reflection coefficients as the features and then implemented SVM as a classifier to get $89\%$ accuracy. Azhiri et al.~\cite{azhiri2021emg} has used the same features but implemented EVM and increased the accuracy to $91\%$. Authors in~\cite{ref} extracted fractional fast Fourier transform (FrFT) as their features, and then KNN was applied as a classifier to achieve the accuracy of $98.12\%$. Bhattachargee et al.~\cite{bhattachargee2019finger} extracted RMS, standard deviation, variance and FFT as their features and classified the results by GB classifier with $98.50\%$ accuracy.

While the majority of studies have concentrated on the offline classification of EMG signals, the online performance of system has practical role on the real-time control of prosthetic hands. For such applications, not only the accuracy of a classifier but also the delay time to get that accuracy is important. Also, the accuracy of offline mode generally is more than online mode. Additionally, the accuracy in online systems depends on the delay time in the postprocessing. The less delay time is considered, the less accuracy is achievable. 

Khushaba et al.~\cite{Khushaba_towards} employed Hudgins features, and autoregressive (AR) for feature extraction and applied library SVM (LIBSVM) for classification. They got the  average accuracy of $90\%$ in $800$ msec. In another online classification approach~\cite{chu2006real}, wavelet packet transform as a generalized format of wavelet transform has been implemented on the EMG signals for feature extraction. Then by employing principal component analysis (PCA) and self-organizing feature map (SOFM), the number of features has been reduced. Finally, a multilayer perceptron (MLP) was used as a classifier. Jaramillo et al.~\cite{jaramillo2017real} applied filtering and rectifying on original signals, extracted features in time, frequency and time-frequency domains, and utilized various parametric and nonparametric classifiers.

 Recently, multifarious deep-learning based classifiers have been utilized in EMG-based hand gesture classification in online mode. Deep learning has the advantage of stability which reduces the errors associated with environmental noises especially in real-time EMG signal classification~\cite{xiong2021deep_review}.
In classification with recurrent neural network (RNN), Nasri et al~\cite{nasri2019inferring_RNN} achieved the accuracy of $77.85\%$ on EMG dataset including $6$ gestures with RNN structure. In order to enhance the accuracy of classification, Koch et el.~\cite{koch2018recurrent_RNN} introduced a new loss function on the outputs of RNN in which true predictions have more weights. The accuracy for true predictions has improved by $10\%$. Reference~\cite{simao2019emg_RNN} compared different structures including feed-forward neural network (FFNN) as a static model and RNN, long-short term memory (LSTM) and gated recurrent units (GRU) as dynamic models to examine the temporal information of EMG signals. Using gesture detection accuracy as a criteria for the evaluation of proposed model, they concluded that both static and dynamic models have similar accuracies.
Also, it seems using other techniques like ensemble learning~\cite{forouzandeh2021presentation} and other neural network architectures such as graph convolutional neural network (GCNN)~\cite{2020new}, to make connection between different sensors, can significantly improve the classification accuracy.

In this paper, we investigate recurrent neural network-based architectures for real-time classification of EMG signals. Discrete wavelet decomposition is employed to extract the features from time-frequency domain to feed into the recurrent neural networks. These architectures achieve a better classification accuracy in using shorter time in comparison with the state-of-the-art methods.

\section{Proposed Architecture}
Our proposed architecture, as shown in Fig.~\ref{Teaser}, includes three main steps: preprocessing, processing, and postprocessing.

\begin{figure}[h]
\centerline{\includegraphics[width=0.4\textwidth]{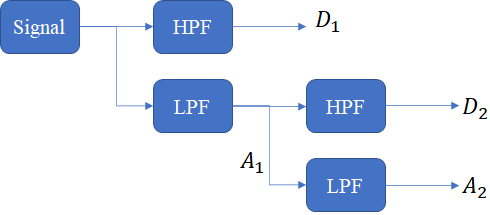}}
\caption{Two-level wavelet Decomposition.}
\label{wavelet}
\end{figure}

\begin{figure}[h]
\centering
\includegraphics[width=0.45\textwidth]{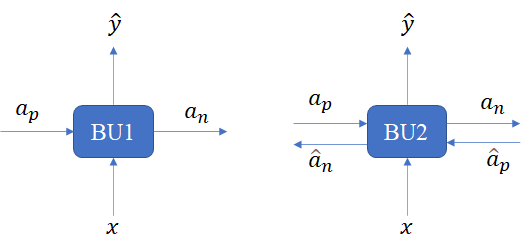}
\caption{Basic units BU1 and BU2.}
\label{BU12}
\end{figure}

\subsection{Preprocessing}
EMG signals have a stochastic behavior such that instantaneous processing is unable to generate favorite information for real-time classification of EMG signals. The input data is divided into the batches of consecutive samples called data windowing. Each window has the size of $400$ samples with $200$ overlaped samples with the previous window. Overlapped windows generally have better classification results than disjointed ones. The wavelet transform features are calculated for each window and used as inputs to the RNN structure. In this paper, we consider a 2-level \textit{db1} mother wavelet transform decomposition.
\begin{figure*} 
\centering
\includegraphics[width=0.7\textwidth]{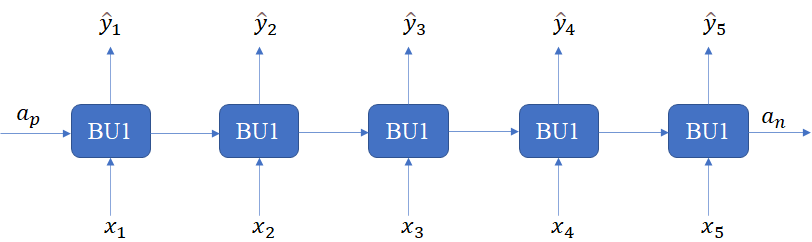}
\caption{The recurrent neural network architecture used in this paper.}
\label{arc1}
\end{figure*}

\begin{figure*} 
\centering
\includegraphics[width=0.7\textwidth]{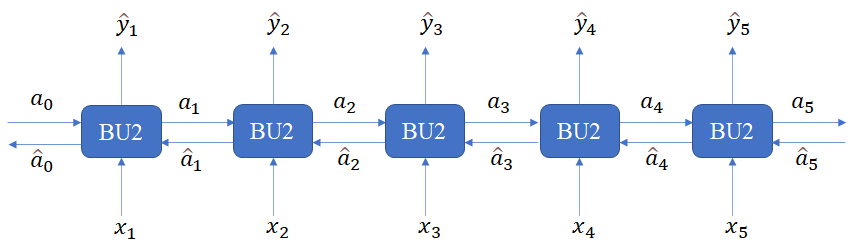}
\caption{The bidirectional recurrent neural network (BRNN) architecture used in this paper.}
\label{arc2}
\end{figure*}

\begin{figure*} 
\centering
\includegraphics[width=0.75\textwidth]{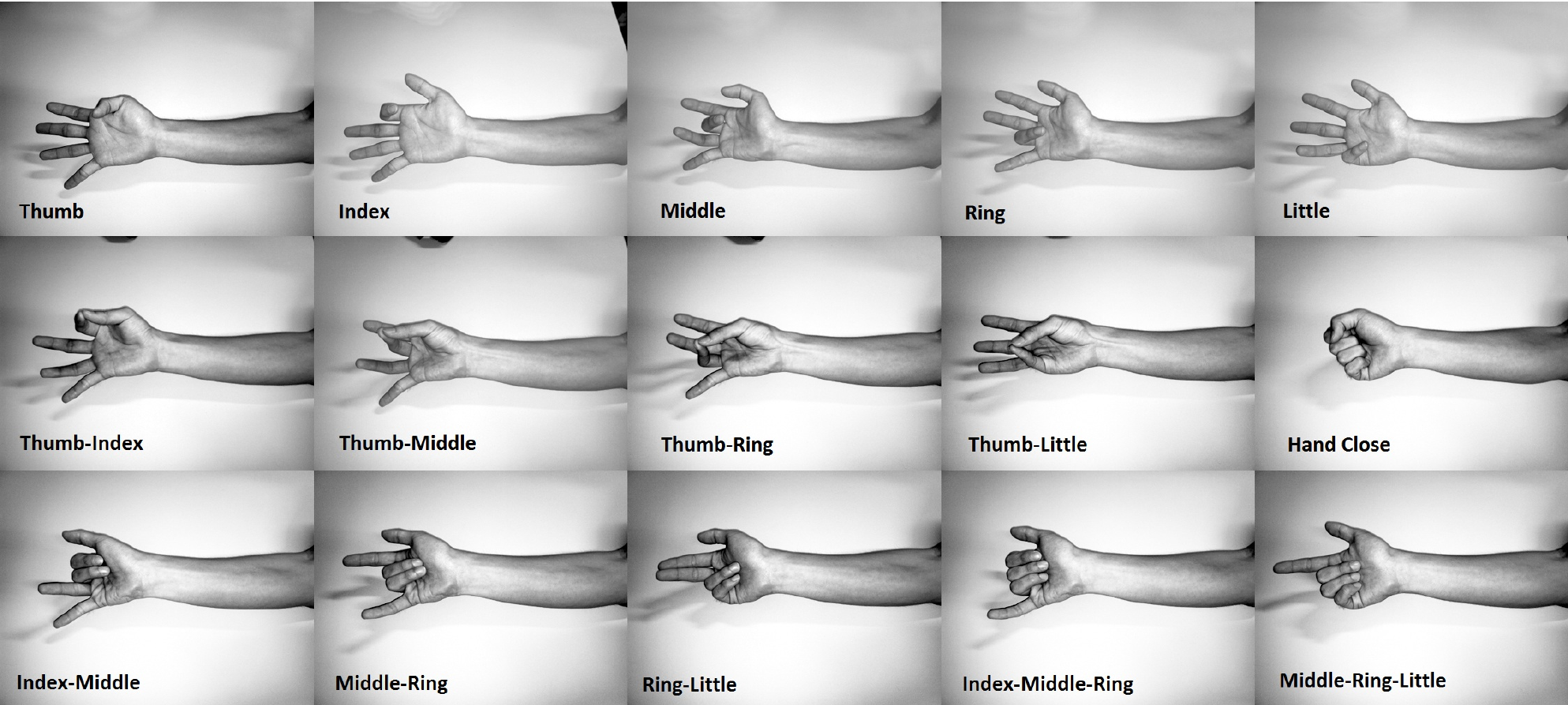}
\caption{Fifteen finger movements classes for individual and combined fingers\cite{khushaba2012electromyogram} Each finger movement has related to a class as: (1. Thumb, 2. Index, 3. Middle, 4. Ring, 5. Little, 6. Thumb-Index, 7. Thumb-Middle, 8. Thumb-Ring, 9. Thumb-Little, 10. Index-Middle (I-M),  11. Middle-Ring (M-R), 12. Ring-Little (RL), 13. Index-Middle-Ring (I-M-R), 14. Middle-Ring-Little (M-RL), and 15. close hands (HC). )}
\label{gestures_15 Class}
\end{figure*}

\subsection{Recurrent Neural Network Structures}
Recurrent neural networks (RNNs) are part of artificial neural networks that are appropriate to exhibit temporal dynamic behavior. RNNs are able to use their memory to process variable length sequences of inputs. This feature makes RNNs a valuable competitor against NNs and CNNs. RNNs have applications such as speech recognition and handwriting recognition, in dealing with sequence data. 

Now we introduce two basic units (BUs) that are employed in our architectures. Fig.~\ref{BU12} illustrates two basic units BU1 and BU2. 
\begin{itemize}
    \item BU1: consider $W_i, b_i$ and $w_a$ as the parameters of the basic unit 1.
    This basic unit takes vectors $x$ and $a_{p}$ as its inputs and returns two outputs including $\hat{y} = A_3$ and $a_n = Z_3$ where
    \begin{align*}
        &A_0 = x \\
        &Z_{i+1} = W_i A_i + b_i \quad \text{for} \quad i = 0,1,2\\
        &A_{i+1} = \text{tanh}(Z_{i+1} + w_{a}a_{p}) \quad \text{for} \quad i = 0 \\ 
         &A_{i+1} = \text{tanh}(Z_{i+1}) \quad \text{for} \quad i = 1 \\ 
        &A_{i+1} = \text{Softmax}(Z_{i+1}) \quad \text{for} \quad i = 2 .
    \end{align*}
    
    \item BU2: consider $W_i, b_i, w_a$ and $w_{\hat{a}}$ as the parameters of the basic unit 1.
    This basic unit takes vectors $x, a_{p}$ and $\hat{p}_{p}$ as its inputs and returns three outputs including $\hat{y} = A_3$ and $a_n = a_p = Z_3$ where
    \begin{align*}
        &A_{i+1} = \text{tanh}(Z_{i+1} + w_{a}a_{p} + w_{\hat{a}}\hat{a}_{p}) \quad \text{for} \quad i = 0 .
    \end{align*}
\end{itemize}

We construct an original recurrent neural network (RNN) and a bidirectional recurrent neural network (BRNN) by stacking a set of the basic units BU1 and BU2,  respectively. The number of basic units that are used at each architecture is a hyperparameter which is tuned empirically. In this paper,  five basic units are employed for each architecture. Also, we consider two different input approaches for each architecture as follows: 
\begin{itemize}
    \item Same inputs: in this approach, all the basic unit inputs are the same, i.e., 
    \begin{align*}
        x_1 = x_2 = x_3 = x_4 = x_5 =  f(w_i)
    \end{align*}
    where $f(w_i)$ refers to the extracted features of window $w_i$ from the original signal. In this method, since all the basic unit inputs are the same, the RNN structure tries to improve the uncertainty at each basic unit during the training phase. 
    \item Sequential inputs: in this approach, the basic unit inputs are different. Indeed, the input of each basic unit is the extracted features from a shifted window (50 msec) of the original signal, i.e., 
    \begin{align*}
        &x_j =  f(w_{i+j-1}) \quad \text{for} \quad j= 1,\cdots, 5 .
    \end{align*}
    In this method, since the basic unit inputs are different, the RNN structure tries to learn the relationship between the inputs to reduce the uncertainty during the training phase~\cite{rouhani2020novel}. 
\end{itemize}

\subsection{Postprocessing}
To refine the results of RNN structure, a postprocessing technique after classification of RNN structure is necessary. Each window performs a class decision, and postprocessing step minimizes ambiguities and false misclassifications~\cite{Khushaba_towards}. In fact, postprocessing step helps us to combine the results gained by each window of the EMG signal. In this paper, the majority voting approach is selected in which the elements of each class is counted and a class as identified by majority is chosen.

\begin{table*}[t]
\centering
\caption{Feature extraction functions used in this paper.}
\begin{tabular}{@{}lcc@{}}
\toprule
Statistical Metric                                                                              & Abbreviation & Formula                                                   \\ \midrule \midrule
Integrated EMG~\cite{zardoshti1995emg}                                         & IEMG         & $\sum_{n=1}^{N} |x[n]|$                                 \\ \midrule
Mean Absolute Value~\cite{hudgins1993new}                                & MAV          & $\frac{1}{N} \sum_{n=1}^{N} |x[n]|$                       \\ \midrule
Simple Square Integrated~\cite{phinyomark2012feature}                      & SSI          & $\sum_{n=1}^{N} x[n]^2$                                   \\ \midrule
Root Mean Square~\cite{phinyomark2012feature}                           & RMS          & $\sqrt{ \frac{1}{N} \sum_{n=1}^{N} x[n]^2 }$              \\ \midrule
Variance~\cite{phinyomark2014feature}                                     & VAR          & $ \frac{1}{N-1} \sum_{n=1}^{N} x[n]^2$                    \\ \midrule
Myopulse Percentage Rate~\cite{phinyomark2012feature}                        & MYOP         & $\frac{1}{N} \sum_{n=1}^{N} f(|x[n]|), \quad f(a) = \left\{\begin{matrix}        1 \quad \text{if} \quad a>T \\         0 \quad \text{otherwise} 
        \end{matrix}\right. $                    \\ \midrule
Waveform Length~\cite{phinyomark2012feature}                             & WL           & $\sum_{n=1}^{N-1} |x[n+1]-x[n]|$                          \\ \midrule
Difference Absolute Mean Value~\cite{phinyomark2014feature}             & DAMV         & $\frac{1}{N-1} \sum_{n=1}^{N-1} |x[n+1]-x[n]|$            \\ \midrule
Second-Order Moment~\cite{phinyomark2014feature}                         & M2           & $\sum_{n=1}^{N-1} (x[n+1]-x[n])^2$                        \\ \midrule
Difference Variance Version~\cite{phinyomark2014feature}                    & DVARV        & $\frac{1}{N-2} \sum_{n=1}^{N-1} (x[n+1]-x[n])^2$          \\ \midrule
Difference absolute standard deviation value~\cite{phinyomark2014feature} & DASDV        & $\sqrt{ \frac{1}{N-1} \sum_{n=1}^{N-1} (x[n+1]-x[n])^2 }$ \\ \midrule
 Maximum~\cite{al2019feature}      & MAX &   $\max x[n]$ \\ \midrule 
 Minimum~\cite{al2019feature}  & MIN        & $\min x[n]$  \\ \midrule 
Willison Amplitude~\cite{phinyomark2012feature}                          & WAMP         & $\sum_{n=1}^{N-1} f(|x[n+1]-x[n]|), \quad f(a) = \left\{\begin{matrix}        1 \quad \text{if} \quad a>T \\         0 \quad \text{otherwise} 
        \end{matrix}\right. $  \\ \midrule
Integrated Absolute of Second Derivative~\cite{azhiri2021emg-based}   & IASD  &  $\sum_{n=1}^{N-2} |x'[n+1]-x'[n]|, \quad x'[n] = x[n+1]-x[n]$ \\ \midrule 
Integrated Absolute of Third Derivative~\cite{azhiri2021emg-based} & IATD &  $\sum_{n=1}^{N-3} |x''[n+1]-x''[n]|, \quad x''[n] = x'[n+1]-x'[n]$  \\ \midrule
Integrated Exponential of Absolute Values\cite{azhiri2021emg-based} & IEAV & $\sum_{n=1}^{N} \exp(|x[n]|)$ \\ \midrule
Integrated Absolute Log Values~\cite{azhiri2021emg-based}  & IALV & $\sum_{n=1}^{N} |\log(x[n] + T)|$ \\ \midrule 
Integrated Exponential~\cite{azhiri2021emg-based} & IE & $ \sum_{n=1}^{N} \exp(x[n])$   \\ \bottomrule
\end{tabular}
\label{conventional features}
\end{table*}

\section{Feature Extraction}
In the time domain, feature extraction functions are directly applied on a window of raw signals. Feature extraction functions are able to extract statistical features from the raw signal. There are several features that are commonly used in the literature.  In this paper, we will use $19$ of these features and their corresponding functions. Table~\ref{conventional features} summarizes the  mathematical expressions of these functions.

Discrete wavelet decomposition (DWD) provides sufficient information both for analysis and synthesis of the original signal, with a significant reduction in the computation time.
Discrete wavelet transform can be implemented with a single level or multiple levels. A two-level discrete wavelet decomposition is shown in
Fig. \ref{wavelet}.

The discrete wavelet transform of a signal is calculated by passing it through a series of filters. The samples are passed through a low pass filter. The signal is also decomposed simultaneously using a high-pass filter. The outputs give the detail coefficients (from the high-pass filter) and approximation coefficients (from the low-pass filter). It is important to note that these two filters are related to each other and they are known as a quadrature mirror filter.
The decomposition is repeated to further increase the frequency resolution and the approximation coefficients decomposed with the high-pass and the low-pass filters and then down-sampled.
After passing a window of raw signal through the high-pass and low-pass filters, the feature extraction functions in Table~\ref{conventional features} are applied on the detail coefficients and approximation coefficients. In this paper, we consider a 2-level \textit{db1} mother wavelet transform decomposition. Therefore, for a two-level wavelet decomposition, we will extract $57$ features ($3$ layers of $19$ wavelet features) for each window of the raw signal. 

\section{Experimental Results}

\subsection{Dataset}

The datasets which are used in this paper are provided by Center of Intelligent Mechatronic Systems at the University of Technology at Sydney\cite{Khushaba_towards}. The datasets includes EMG data of eight participants (six men and two women in range of $20$ to $30$ years old). EMG signals are sensitive to the condition of experiment and effects of neighbor limbs. For uniformity, participants sat on the armchair such that their arms were fixed and stable. To firmly stick the sensors to the skin above the targeted muscle, an adhesive skin interface was used. Resulting EMG signals from the electrodes were amplified to get the total gain of $1000$. The data was recorded at $4,000$ Hz applying an analog-to-digital converter, a bandpass filter between $20$ and $40$ Hz and a notch filter to remove the $50$ Hz line interference. This filtering is necessary to remove noises resulting by motion of artifacts and high-frequency random noise. Participants performed each movement of fingers six times while the duration of each movement from rest pose to contraction pose is five seconds.

In the first dataset, participants performed ten different finger movements consisting five individuals shown as thumb (T), index (I), middle (M), ring (R), little (L) and five combined movements named as thumb-index (T-I), thumb-middle (T-M), thumb-ring(T-R), thumb-little (T-L) and closed hands using two-channel (2C) sensors. We select the first four trials for the training of the classification methods and the last two for the test of the methods in order to check the accuracy of the classifiers. Each gesture took five seconds including rest and holding of each finger posture. 

Second dataset includes total of $15$ classes: five individuals finger movement thumb (T), index (I), middle (M), ring (R), little (L) and ten combined finger movements including thumb-index (T-I), thumb-middle (T-M), thumb-ring (T-R), thumb-little (TL), index-middle (I-M), middle-ring (M-R), ring-little (RL), index-middle-ring (I-M-R), middle-ring-little (M-RL), and close hands (HC) using eight-channel (8C) sensors. Each gesture took 20 seconds including rest and holding of each finger gesture. The first two trials were selected for the training of the model and the last one was used for the test to check the accuracy of proposed classifiers. Fig.~\ref{gestures_15 Class} depicts these datasets that first ten finger movements are considered in the first dataset and entire fifteen movements are considered in the second dataset. The details of dataset, number of subjects and condition of experiments are reported in Table~\ref{Specifications of EMG datasets}.

\begin{table}[t]
\centering
\caption{Specifications of EMG datasets. }
\begin{tabular}{lll}
\toprule
Specifications                        & \multicolumn{1}{l}{2C Dataset}  & \multicolumn{1}{l}{8C Dataset}                                              \\ \midrule \midrule
Number of electrodes                  & 2                        & 8                                        \\
Number of finger movements            & 10                        & 15                                         \\
Number of repititions                 & 3                         & 3                                            \\
Total number of repitions per subject & 30                       & 45                                            \\
Number of subjects                    & 8               & 8                                                \\
Total number of repititions           & 360                 & 360                                         \\
Time for each repitition              & 5 sec             & 20 sec                                            \\
Sampling rate                         & 4000 Hz            & 4000 Hz                                         \\
Resolutions                           & 12 bit         & 12 bit                                                  \\
Window length/overlap                & 100/50 msec & 100/50 msec  \\ \bottomrule
\end{tabular}
\label{Specifications of EMG datasets}
\end{table}

\begin{table*}[ht]
\centering
\caption{The accuracy (in \%) of different RNN structures for various signal length for 2C dataset.}
\begin{tabular}{@{}lccccccccccc@{}}
\toprule
\multicolumn{1}{c}{\multirow{2}{*}{RNN Structures}} & \multicolumn{11}{c}{Signal length (msec)}                                    \\ \cmidrule(l){2-12} 
\multicolumn{1}{c}{}                                & 100  & 150  & 200  & 250  & 300  & 350  & 400  & 450  & 500  & 550  & 600  \\ \midrule  \midrule
RNN with the same inputs                                                 & 62.0   & 62.0   & 72.5 & 74.0   & 78.0   & 79.5 & 84.5 & 86.0   & 85.5 & 89.0   & 90.0   \\ \midrule
BRNN with the same inputs                                   & 70.5 & 70.5 & 74.0   & 81.5 & 87.0   & 90.0   & 91.0   & 92.5 & 92.0   & 91.5 & 96.0   \\ \midrule
RNN with sequential inputs& -    & -    & -    & -    & 85.5 & 85.5 & 88.0   & 89.0   & 91.0   & 93.0   & 93.0 
 \\ \midrule
 BRNN with sequential inputs                                         & -    & -    & -    & -    & 85.0   & 85.0   & 91.0   & 91.5 & 93.5 & 93.5 & 93.5
  \\ \bottomrule
\end{tabular}
\label{RNN for two channel}
\end{table*}

\begin{table*}[ht]
\centering
\caption{The accuracy (in \%) of different RNN structures for various signal length for 8C dataset.}
\begin{tabular}{@{}lccccccccccc@{}}
\toprule
\multicolumn{1}{c}{\multirow{2}{*}{RNN Structures}} & \multicolumn{11}{c}{Signal length (msec)}                                    \\ \cmidrule(l){2-12} 
\multicolumn{1}{c}{}                                & 100  & 150  & 200  & 250  & 300  & 350  & 400  & 450  & 500  & 550  & 600  \\ \midrule  \midrule
RNN with the same inputs                                                  & 82.5 & 82.5 & 81.6 & 84.1 & 87.5 & 84.1 & 86.6 & 87.5 & 90.8 & 91.6 & 91.6   \\ \midrule
BRNN with the same inputs                                   & 85.0 & 85.0 & 90.0 & 90.8 & 92.5 & 92.5 & 92.5 & 92.5 & 93.3 & 93.3 & 93.3   \\ \midrule
RNN with sequential inputs                            & -    & -    & -    & -    & 89.1 & 89.1 & 90.8 & 90.8 & 90.8 & 91.6 & 92.5 
\\ \midrule
BRNN with sequential inputs                                        & -    & -    & -    & -    & 89.2 & 89.2 & 90.8 & 90.8 & 93.3 & 93.3 & 93.3 
  \\ \bottomrule
\end{tabular}
\label{RNN for eight channel}
\end{table*}

\begin{table*}[ht]
\centering
\caption{COMPARING THE RESULTS OF DIFFERENT APPROACHES FOR THE SAME (2C) EMG DATASET~\cite{Khushaba_towards}.}
\begin{tabular}{ >{\centering\arraybackslash}p{1.10cm} >{\centering\arraybackslash}p{3.30cm} >{\centering\arraybackslash}p{2.65cm} >{\centering\arraybackslash}p{1.25cm}>{\centering\arraybackslash}p{3.25cm}>{\centering\arraybackslash}p{3.50cm} }
  \toprule
Methods & Features & Classifier &  Classes & Signal length (msec) & Avg. Acc (in \%) \\
  \midrule
  \midrule
  \cite{ariyanto2015finger} & TD+Hjorth+RMS & ANN & 5 & 5000 & 96.7 \\
  \midrule
  \cite{naik2014nonnegative} & AR + RMS & ANN+NMF & 5 & 5000 & 92 \\
  \midrule
  \cite{Heydarzade} & Reflection Coefficients & SVM & 10 & 5000 & 89 \\
  \midrule
  \cite{khushaba2011electromyogram} & FNPA & ELM+libSVM+RegTree & 10 & 5000 & 91 \\
  \midrule
  \cite{azhiri2021emg} & Reflection Coefficients & EVM & 10 & 5000 & 91 \\
  \midrule
  \cite{al2019enhancing} & Mixture of Features & Random Forest & 10 & 5000 & 93.75 \\
  \midrule
  \cite{esa2018electromyography} & Hudgind+RMS & SVM & 10 & 5000 & 96.67 (just thumb finger)\\
  \midrule
  \cite{taghizadeh2021finger} & FrFT & KNN & 10 & 5000 & 98.12 \\
  \midrule 
  \cite{bhattachargee2019finger} & statistics features+FFT & GB & 10 & 5000 & 98.5 \\
  \midrule
  \cite{phukan2019finger} & Mixture of Features & SVM & 10 & 5000 & 96.5\\
  \midrule
  \cite{Khushaba_towards} & TD+AR+Hjorth & KNN+SVM+Fusion & 10  &  800 & 90  \\
  \midrule
  \cite{azhiri2021emg} & Wavelet & Deep Neural Network & 10 &  800  & 95.5 \\
  \midrule
  \textbf{Ours} & \textbf{Wavelet} & \textbf{BRNN} & \textbf{10} & \textbf{ 600} & \textbf{96}  \\
  \bottomrule
\end{tabular}
\label{comparing two channel}
\end{table*}

\begin{table*}[ht]
  \centering
\caption{COMPARING THE RESULTS OF DIFFERENT APPROACHES FOR THE SAME (8C) EMG DATASET~\cite{khushaba2012electromyogram}.}
\begin{tabular}{ >{\centering\arraybackslash}p{1.10cm} >{\centering\arraybackslash}p{3.30cm} >{\centering\arraybackslash}p{2.65cm} >{\centering\arraybackslash}p{1.25cm}>{\centering\arraybackslash}p{3.25cm}>{\centering\arraybackslash}p{3.50cm} }
  \toprule
Methods & Features & Classifier &  Classes & Signal length (msec) & Avg. Acc (in \%) \\
  \midrule
  \midrule
  \cite{khushaba2012electromyogram} & MCA & SVM+KNN+ELM & 15 & 20000 & 95.0 \\
  \midrule
  \cite{shin2014performance} & Mixture of Features & NB+KNN+MLP+
  QDA+SVM+ELM & 15 & 20000 & 90 \\
  \midrule
  \cite{bhagwat2020electromyogram} & WPT-4+TD+AR-4+RMS & QDA+KNN+SVM  & 15 & 20000 & 98.5 \\
  \midrule
  \cite{jafarzadeh2019deep} & Raw Data & CNN & 15 & 100 & 91.26 \\
  \midrule
  \textbf{Ours} & \textbf{Wavelet} & \textbf{BRNN} & \textbf{15} & \textbf{500} & \textbf{93.33}  \\
  \bottomrule
\end{tabular}
\label{comparing eight channel}
\end{table*}

\begin{figure*}  
\centering
\includegraphics[width=0.6\textwidth]{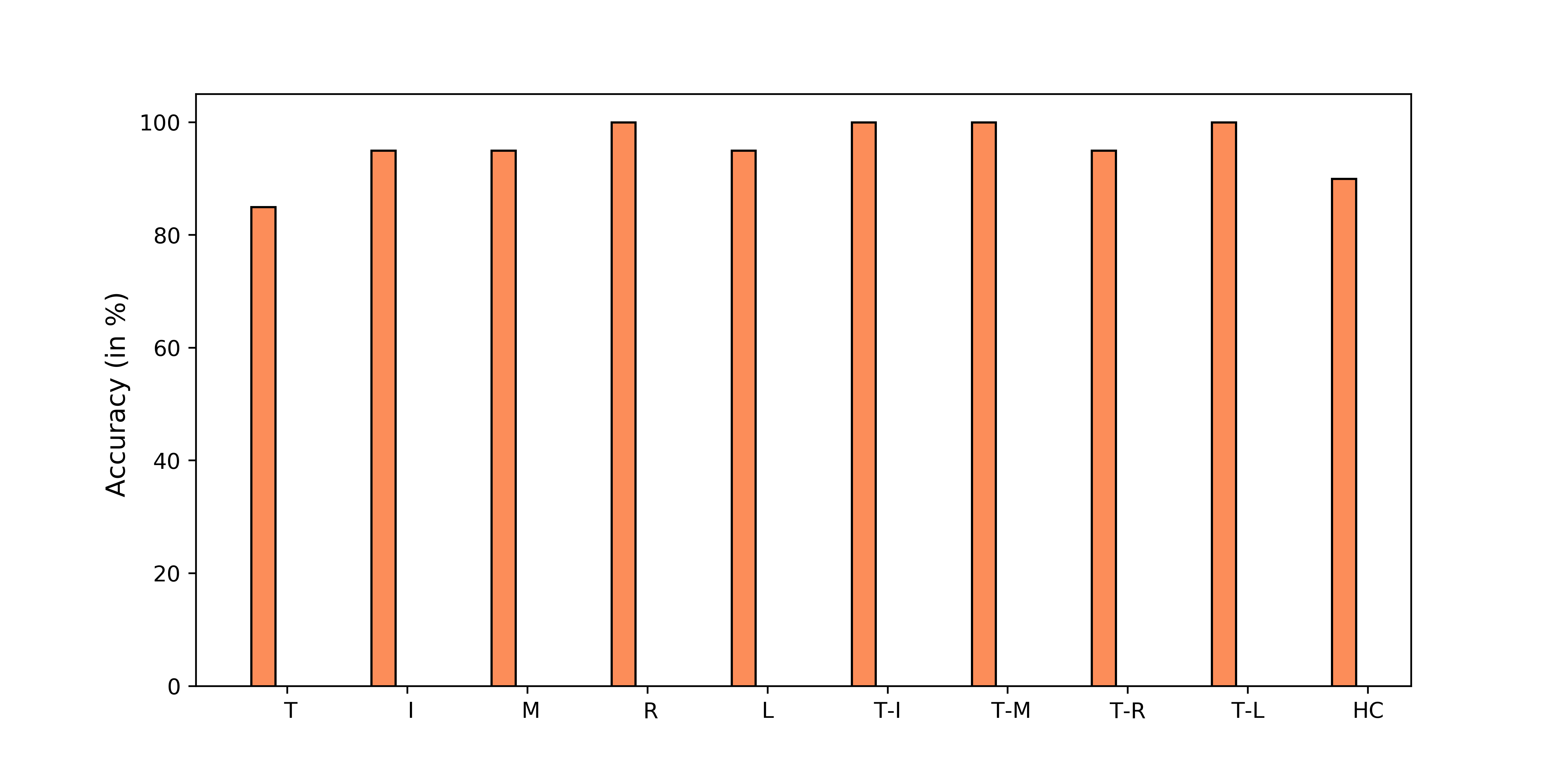}
\caption{The classification accuracy for each of $10$ classes of BRNN for 2C dataset at signal length of 600 msec.}
\label{each_class_accuracy_2}
\end{figure*}

\begin{figure*}  
\centering
\includegraphics[width=0.6\textwidth]{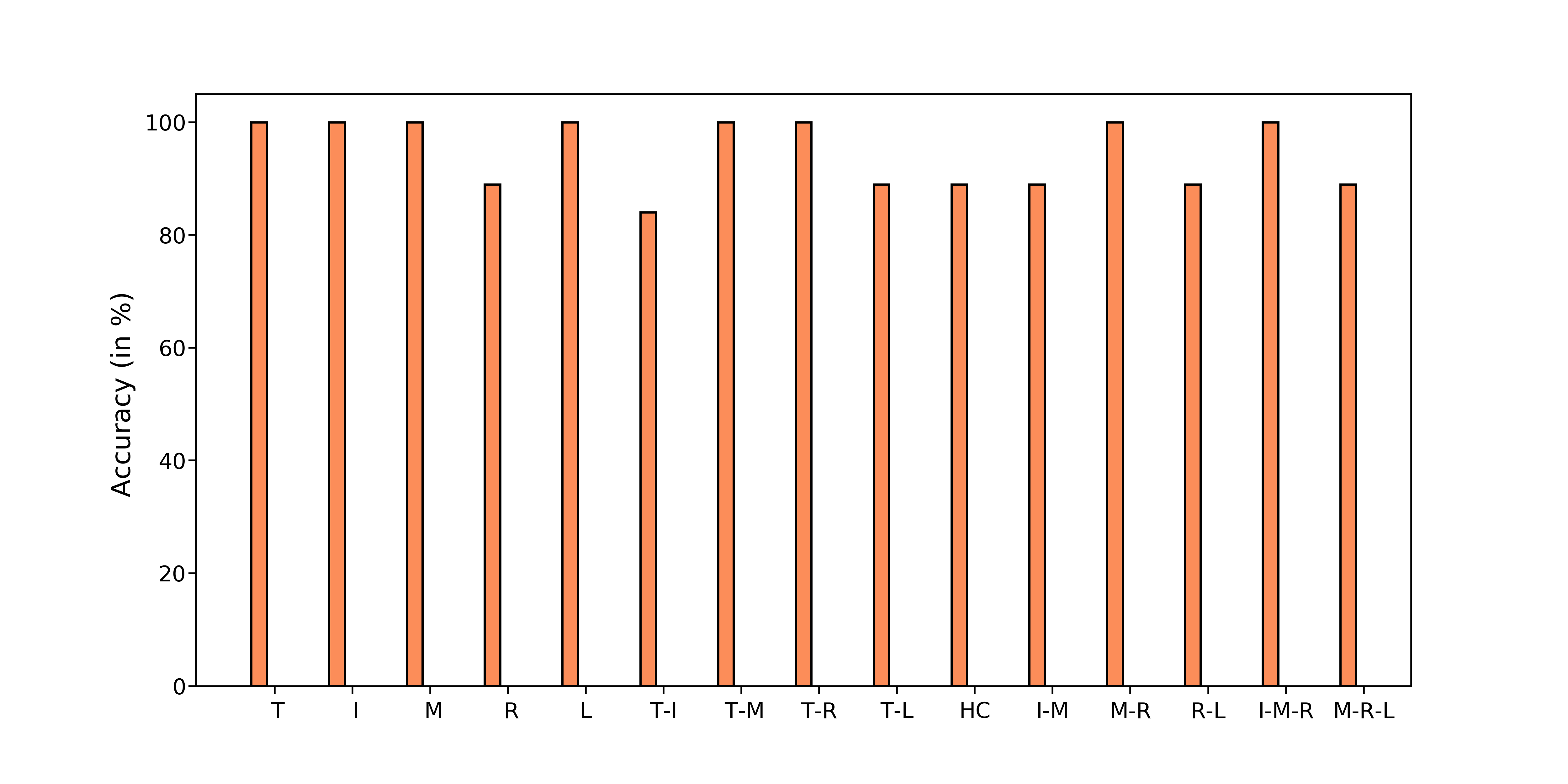}
\caption{The classification accuracy for each of $15$ classes of BRNN for 8C dataset at signal length of 500 msec.}
\label{each_class_accuracy_8}
\end{figure*}

\subsection{Results and Discussion}
In this section, the classification results of RNN and bidirectional RNN (BRNN) are presented and compared. Table~\ref{RNN for two channel} compares the test accuracy of proposed RNN architectures at different signal lengths for the first dataset. The method which is used for the postprocessing is majority voting. The results show that BRNN with the same inputs could achieve the higher accuracy in $600$ msec which is the best performance among all other proposed architectures of RNN. 
Due to the two-sided connectivity of basic units, the BRNN architecture is able to capture the relationship of inputs better than a one-sided RNN architecture. Indeed, sharing the features that are learned by each basic unit reduces the misclassification and improves overall accuracy. Also, considering the same inputs for all basic units enhances the classifier performance by providing a prior information about the inputs. In other words, the uncertainty that exist between the basic units is reduced in BRNN compared to RNN.

Again, various architectures of RNN at different signal lengths are compared for eight-channel (8C) dataset. BRNN  with the same inputs and RNN with the sequential inputs show better performance than other architectures. Both architectures achieve the accuracy of $93.3\%$ in $500$ msec. These results are summarized in Table~\ref{RNN for eight channel}.  

In Fig.~\ref{each_class_accuracy_2} and Fig.~\ref{each_class_accuracy_8}, the classification accuracy of 2C and 8C datasets for BRNN with the same inputs are shown. For the 2C dataset at the $600$ msec signal length, the accuracy of ring, thumb-index, thumb-middle and thumb-little is $100\%$. For 8C dataset at the $500$ msec signal length, the accuracy for thumb, index, middle, little, thumb-middle, thumb-ring, middle-ring and index-middle-ring is $100\%$. Comparing the accuracies of classes, a uniform accuracy for each finger movement in both datasets is observed. Such uniform accuracy for different classes is important factor that proves the proposed BRNN architecture has stable performance on various finger movements. 

The accuracy of classification depends on several factors including the number of classes, number of sensors, extracted features and utilized classifier. Increasing the number of classes will decrease the accuracy of classifications. Also, the number of sensors used for the experiments can significantly affect the final result. Therefore, we compare our results with other researchers with the same situation on the same dataset. In Table~\ref{comparing two channel}, the results of 2C EMG dataset are compared, while in Table~\ref{comparing eight channel}, the 8C EMG dataset is considered. The offline results have generally better accuracies but as it is discussed, for real world applications and motion control of prosthetics, online classification have practical role. Rare researches have been done on the online classification of EMG signals. Our result is very competitive such that the accuracy and signal length are simultaneously regarded and better accuracy in less signal length is gained for both datasets. 

We have trained this model in python and ran it in a windows-based PC with 2.60 GHz CPU and 16 GB memory. The delay time for the system is the result of feature extraction and process time required by the classifier to make the decision. For 2C dataset using BRNN at $600$ msec signal length at postprocessing, delay time is about $4.72$ msec for the feature extraction and $3.12$ msec for the classification.


\section{Conclusion}


To overcome the challenges of real-time classification of EMG signals, we introduced two new recurrent neural network-based architectures. 
Discrete wavelet transformation was utilized to extract the features from time-frequency domain to feed into the recurrent neural networks.
Each architecture was investigated for two different input forms. The performance of each architecture was evaluated and it was compared with the state-of-the-art approaches.
It was shown that at least one of these  architectures increases the classification accuracy, reduces the delay time, and outperforms other architectures.

\bibliographystyle{ieeetr}
\bibliography{Ref.bib}
\end{document}